\def\year{2021}\relax
\documentclass[letterpaper]{article} 
\usepackage{aaai21}  
\usepackage{times}  
\usepackage{helvet} 
\usepackage{courier}  
\usepackage[hyphens]{url}  
\usepackage{graphicx} 
\usepackage{natbib}  
\usepackage{caption} 
\usepackage{cite}
\usepackage{threeparttable}
\usepackage{amsmath,amssymb,amsfonts}
\usepackage{graphicx}
\usepackage{textcomp}
\usepackage{xcolor}
\usepackage{float}
\usepackage{array}
\usepackage{tabularx}
\usepackage{array}
\usepackage{balance}
\usepackage{marvosym}
\usepackage{multirow}
\usepackage{array}
\usepackage{enumitem}
\usepackage{booktabs}
\usepackage{algpseudocode}
\usepackage[ruled]{algorithm2e} 
\usepackage{url}

\title{Self-Supervised Hypergraph Convolutional Networks for \\Session-based Recommendation}
\author{
    Xin Xia\textsuperscript{\rm 1},
    Hongzhi Yin\textsuperscript{\rm 1}\thanks{Corresponding author},
    Junliang Yu\textsuperscript{\rm 1},
    Qinyong Wang\textsuperscript{\rm 1},
    Lizhen Cui\textsuperscript{\rm 2},
    Xiangliang Zhang\textsuperscript{\rm 3}\\
}
\affiliations{

    \textsuperscript{\rm 1}The University of Queensland,\\ 
    \textsuperscript{\rm 2}Shandong University, \\
    \textsuperscript{\rm 3}King Abdullah University of Science and Technology\\

     \{x.xia, h.yin1, jl.yu, qinyong.wang\}@uq.edu.au, clz@sdu.edu.cn, xiangliang.zhang@kaust.edu.sa

}
\begin{document}

\maketitle

\begin{abstract}
	Session-based recommendation (SBR) focuses on next-item prediction at a certain time point. As user profiles are generally not available in this scenario, capturing the user intent lying in the item transitions plays a pivotal role. Recent graph neural networks (GNNs) based SBR methods regard the item transitions as pairwise relations, which neglect the complex high-order information among items. Hypergraph provides a natural way to capture beyond-pairwise relations, while its potential for SBR has remained unexplored. In this paper, we fill this gap by modeling session-based data as a hypergraph and then propose a hypergraph convolutional network to improve SBR. Moreover, to enhance hypergraph modeling, we devise another graph convolutional network which is based on the line graph of the hypergraph and then integrate self-supervised learning into the training of the networks by maximizing mutual information between the session representations learned via the two networks, serving as an auxiliary task to improve the recommendation task. Since the two types of networks both are based on hypergraph, which can be seen as two channels for hypergraph modeling, we name our model \textbf{DHCN} (Dual Channel Hypergraph Convolutional Networks). Extensive experiments on three benchmark datasets demonstrate the superiority of our model over the SOTA methods, and the results validate the effectiveness of hypergraph modeling and self-supervised task. The implementation of our model is available via \url{https://github.com/xiaxin1998/DHCN}.
\end{abstract}

\section{Introduction}
Session-based recommendation (SBR) is an emerging recommendation paradigm, where long-term user profiles are usually not available~\cite{wang2019survey,guo2019streaming}. Generally, a session is a transaction with multiple purchased items in one shopping event, and SBR focuses on next-item prediction by using the real-time user behaviors. Most of the research efforts in this area regard the sessions as ordered sequences, among which recurrent neural networks (RNNs) based \cite{hidasi2015session,jannach2017recurrent,hidasi2018recurrent} and graph neural networks (GNNs) \cite{wu2020comprehensive} based approaches have shown great performance. 

In RNNs-based approaches, modeling session-based data as unidirectional sequences is deemed as the key to success, since the data is usually generated in a short period of time and is likely to be temporally dependent. However, this assumption may also trap these RNNs-based models because it ignores the coherence of items. Actually, unlike linguistic sequences which are generated in a strictly-ordered way, among user behaviors, there may be no such strict chronological order. For example, on Spotify\footnote{https://www.spotify.com/}, a user can choose to shuffle an album or play it in order, which generates two different listening records. However, both of these two play modes serialize the same set of songs. In other words, reversing the order of two items in this case would not lead to a distortion of user preference. Instead, strictly and solely modeling the relative orders of items and ignoring the coherence of items would probably make the recommendation models prone to overfitting. 

Recently, the effectiveness of graph neural networks (GNNs) \cite{wu2020comprehensive,yu2020enhance,yin2019social} has been reported in many areas including SBR. Unlike the RNNs-based recommendation method, the GNNs-based approaches \cite{wu2019session,xu2019graph,qiu2020exploiting} model session-based data as directed subgraphs and item transitions as pairwise relations, which slightly relaxes the assumption of temporal dependence between consecutive items. However, existing models only show trivial improvements compared with RNNs-based methods. The potential reason is that they neglect the complex item correlations in session-based data. In real scenarios, an item transition is often triggered by the joint effect of previous item clicks, and many-to-many and high-order relations exist among items. Obviously, simple graphs are incapable of depicting such set-like relations. 

\begin{figure*}[t]
	\centering
	\includegraphics[width=\textwidth]{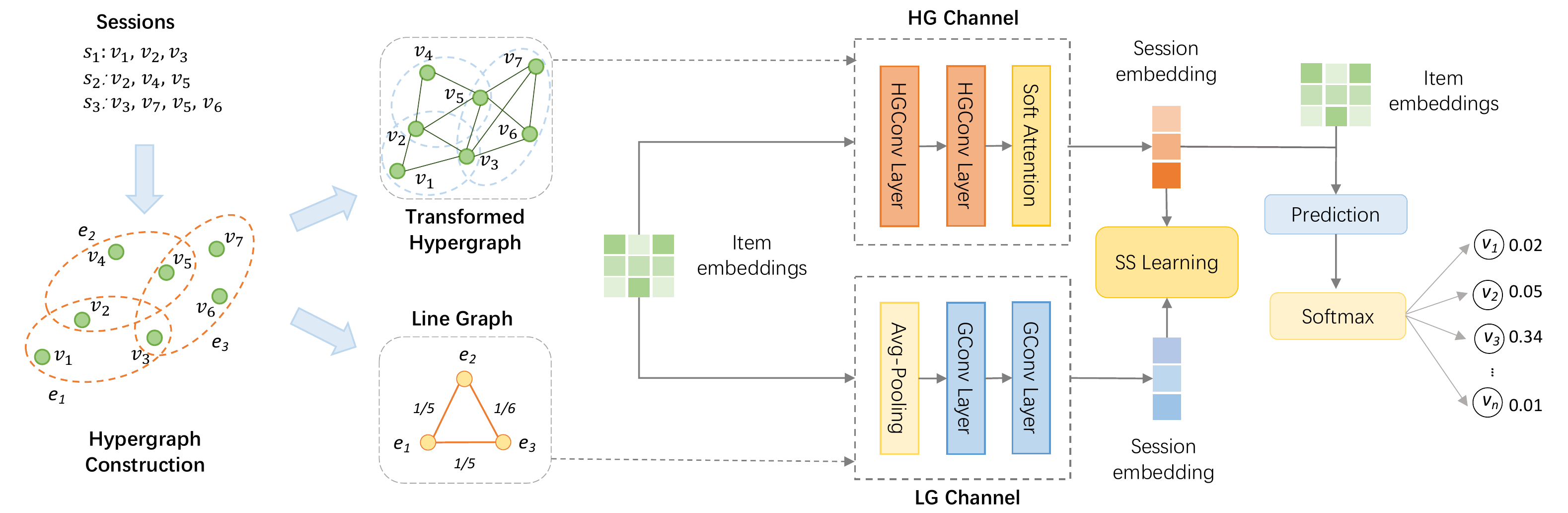}
	\caption{The construction of hypergraph and the pipeline of the proposed DHCN model. }
	\label{figure.1}
	\vspace{-15pt}
\end{figure*}

To overcome these issues, we propose a novel SBR approach upon hypergraph to model the high-order relations among items within sessions. Conceptually, a hypergraph \cite{bretto2013hypergraph} is composed of a vertex set and a hyperedge set, where a hyperedge can connect any numbers of vertices, which can be used to encode high-order data correlations. We also assume that items in a session are temporally correlated but not strictly sequentially dependent. The characteristics of hyperedge perfectly fit our assumption as hyperedge is set-like, which emphasizes coherence of the involved elements rather than relative orders. Therefore, it provides us with a flexibility and capability to capture complex interactions in sessions. Technically, we first model each session as a hyperedge in which all the items are connected with each other, and different hyperedges, which are connected via shared items, constitute the hypergraph that contains the item-level high-order correlations. Figure \ref{figure.1} illustrates the hypergraph construction and the pipeline of the proposed method. 
\par 
By stacking multiple layers in the hypergraph channel, we can borrow the strengths of hypergraph convolution to generate high-quality recommendation results. However, since each hyperedge only contains a limited number of items, the inherent data sparsity issue might limit the benefits brought by hypergraph modeling. To address this problem, we introduce line graph channel and innovatively integrate self-supervised learning \cite{hjelm2018learning} into our model to enhance hypergraph modeling. A line graph is built based on the hypergraph by modeling each hyperedge as a node and focuses on the connectivity of hyperedges, which depicts the session-level relations.  After that, a \textbf{D}ual channel \textbf{H}ypergraph \textbf{C}onvolutional \textbf{N}etwork (DHCN) is developed in this paper with its two channels over the two graphs. Intuitively, the two channels in our network can be seen as two different views that describe the intra- and inter- information of sessions, while each of them knows little information of the other. By maximizing the mutual information between the session representations learned via the two channels through self-supervised learning, the two channels can acquire new information from each other to improve their own performance in item/session feature extraction. We then unify the recommendation task and the self-supervised task under a \textit{primary\&auxiliary} learning framework. By jointly optimizing the two tasks, the performance of the recommendation task achieves decent gains.
 \par
Overall, the main contributions of this work are summarized as follows:
\begin{itemize}
	\item We propose a novel dual channel hypergraph convolutional network for SBR, which can capture the beyond-pairwise relations among items through hypergraph modeling.	
	\item We innovatively integrate a self-supervised task into the training of our network to enhance hypergraph modeling and improve the recommendation task.	
	\item Extensive experiments show that our proposed model has overwhelming superiority over the state-of-the-art baselines and achieves statistically significant improvements on benchmark datasets.
\end{itemize}
%

\section{Related Work}
\label{st:rw}
\subsection{Session-based Recommendation}
The initial exploration of SBR mainly focuses on sequence modeling, where Markov decision process is the preferred technique at this phase. \cite{shani2005mdp,rendle2010factorizing,zimdars2013using} are the representative works of this line of research. The boom of deep learning provides alternatives to exploit sequential data. Deep learning models such as recurrent neural networks \cite{hochreiter1997long,cho2014learning} and convolutional neural networks \cite{tuan20173d} have subsequently been applied to SBR and achieved great success. \cite{hidasi2015session,tan2016improved,li2017neural,liu2018stamp} are the classical RNNs-based models which borrow the strengths of RNNs to model session-based data.\par 
Graph Neural Networks (GNNs) \cite{wu2020comprehensive,zhou2018graph} recently have drawn increasing attention and their applications in SBR also have shown promising results \cite{wang2020beyond, wang2020global,yuan2019simple,chen2020handling}. Unlike RNNs-based approaches working on sequential data, GNNs-based methods learn item transitions over session-induced graphs. SR-GNN \cite{wu2019session} is the pioneering work which uses a gated graph neural network to model sessions as graph-structured data. GC-SAN \cite{xu2019graph} employs self-attention mechanism to capture item dependencies via graph information aggregation. FGNN \cite{qiu2019rethinking} constructs a session graph to learn item transition pattern and rethinks the sequence order of items in SBR. GCE-GNN \cite{wang2020global} conduct graph convolution on both the single session graph and the global session graph to learn session-level and global-level embeddings. Although these studies demonstrate that GNN-based models outperform other approaches including RNNs-based ones, they all fail to capture the complex and higher-order item correlations. 

\subsection{Hypergraph Learning}

Hypergraph provides a natural way to capture complex high-order relations. With the boom of deep learning, hypergraph neural network also have received much attention. HGNN \cite{feng2019hypergraph} and HyperGCN \cite{yadati2019hypergcn} are the first to apply graph convolution to hypergraph. \cite{jiang2019dynamic} proposed a dynamic hypergraph neural network and \cite{bandyopadhyay2020line} developed the line hypergraph convolutional networks.
\par
There are also a few studies combining hypergraph learning with recommender systems \cite{bu2010music,li2013news}. The most relevant work to ours is HyperRec \cite{wang2020next}, which uses hypergraph to model the short-term user preference for next-item recommendation. However, it does not exploit inter-hyperedge information and is not designed for session-based scenarios. Besides, the high complexity of this model makes it impossible to be deployed in real scenarios. Currently, there is no research bridging hypergraph neural networks and SBR, and we are the first to fill this gap.

\subsection{Self-supervised Learning}
Self-supervised learning \cite{hjelm2018learning} is an emerging machine learning paradigm which aims to learn the data representation from the raw data. It was firstly used in visual representation learning \cite{bachman2019learning}. The latest advances in this area extend self-supervised learning to graph representation learning \cite{velickovic2019deep}. The dominant paradigm based on contrastive learning \cite{hassani2020contrastive,qiu2020gcc} suggests that contrasting congruent and incongruent views of graphs with mutual information maximization can help encode rich graph/node representations.
\par
As self-supervised learning is still in its infancy, there are only several studies combining it with recommender systems \cite{zhou2020s,ma2020disentangled,xin2020self}. The most relevant work to ours is S$^{3}$-Rec \cite{zhou2020s} for sequential recommendation, which uses feature mask to create self-supervision signals. But it is not applicable to SBR since the session data is very sparse and masking features cannot generate strong self-supervision signals. Currently, the potentials of self-supervised learning for hypergraph representation learning and SBR have not been investigated. We are the first to integrate self-supervised learning into the scenarios of SBR and hypergraph modeling.

\section{The Proposed Method}
\label{st:method}
In this section, we first introduce the notions and definitions used throughout this paper, and then we show how session-based data is modeled as a hypergraph. After that, we present our hypergraph convolutional network for SBR. Finally, we devise the line graph channel and integrate self-supervised learning into the dual channel network to enhance hypergraph modeling. 
\subsection{Notations and Definitions}
Let $I = \{i_{1}, i_{2}, i_{3}, ... , i_{N}\}$ denote the set of items, where $N$ is the number of items. Each session is represented as a set $s = [i_{s,1}, i_{s,2}, i_{s,3}, ... , i_{s,m}]$ and $i_{s,k}\in I (1 \leq k \leq m)$ represents an interacted item of an anonymous user within the session $s$. We embed each item $i\in I$ into the same space and let $\mathbf{x}_{i}^{(l)} \in \mathbb{R}^{d^{(l)}}$ denote the vector representation of item $i$ of dimension $d^{l}$ in the $l$-th layer of a deep neural network. The representation of the whole item set is denoted as $\mathbf{X}^{(l)}\in\mathbb{R}^{N \times d^{(l)}}$. Each session $s$ is represented by a vector $\mathbf{s}$. The task of SBR is to predict the next item, namely $i_{s,m+1}$, for any given session $s$. 

\noindent\textbf{Definition 1. Hypergraph.} Let $G= (V,E)$ denote a hypergraph, where $V$ is a set containing $N$ unique vertices and $E$ is a set containing $M$ hyperedges. Each hyperedge $\epsilon \in E$ contains two or more vertices and is assigned a positive weight $W_{\epsilon\epsilon}$, and all the weights formulate a diagonal matrix $\mathbf{W} \in \mathbb{R}^{M \times M}$. The hypergraph can be represented by an incidence matrix $\mathbf{H} \in \mathbb{R}^{N \times M}$ where $H_{i \epsilon}$ = 1 if the hyperedge $\epsilon \in E$ contains a vertex $v_i \in V$, otherwise 0. For each vertex and hyperedge, their degree $D_{i i}$ and $B_{\epsilon \epsilon}$ are respectively defined as $D_{i i}=\sum_{\epsilon=1}^{M} W_{\epsilon \epsilon} H_{i \epsilon}; B_{\epsilon\epsilon}=\sum_{i=1}^{N} H_{i \epsilon}$. $\mathbf{D}$ and $\mathbf{B}$ are diagonal matrices.\par
\noindent\textbf{Definition 2. Line graph of hypergraph.} Given the hypergraph $G = (V,E)$, the line graph of the hypergraph $L(G)$ is a graph where each node of $L(G)$ is a hyperedge in $G$ and two nodes of $L(G)$ are connected if their corresponding hyperedges in $G$ share at least one common node \cite{whitney1992congruent}. Formally, $L(G) = (V_{L}, E_{L})$ where $V_{L} = \{v_{e}: v_{e} \in E\}$, and $E_{L} = \{(v_{e_{p}}$, $v_{e_{q}}): e_{p}$, $e_{q}\in E, |e_{p}\cap e_{q}| \geq 1\}$. We assign each edge $(v_{e_{p}}$, $v_{e_{q}})$ a weight $W_{p,q}$, where 
$W_{p, q}=|e_{p} \cap e_{q}|/|e_{p} \cup e_{q}|$.

\subsection{Hypergraph Construction}
To capture the beyond pairwise relations in session-based recommendation, we adopt a hypergraph $G = (V,E)$ to represent each session as a hyperedge. Formally, we denote each hyperedge as $[i_{s,1}, i_{s,2}, i_{s,3}, ... , i_{s,m}] \in E$ and each item $i_{s,m}\in V$. The changes of data structure before and after hypergraph construction are shown in the left part of Figure \ref{figure.1}. As illustrated, the original session data is organized as linear sequences where two items $i_{s,m-1}, i_{s,m}$ are connected only if a user interacted with item $i_{s,m-1}$ before item $i_{s,m}$. After transforming the session data into a hypergraph, any two items clicked in a session are connected. It should be noted that we transform the session sequences into an undirected graph, which is in line with our intuition that items in a session are temporally related instead of sequentially dependent. By doing so, we manage to concretize the many-to-many high-order relations. Besides, we further induce the line graph of the hypergraph according to Definition 2. Each session is modeled as a node and different sessions are connected via shared items. Compared with the hypergraph which depicts the item-level high-order relations, the line graph describes the session-level relations that are also termed cross-session information. \par

\subsection{Hypergraph Convolutional Network}
After the hypergraph construction, we develop a hypergraph convolutional network to capture both the item-level high-order relations. 
\subsubsection{Hypergraph Channel and Convolution}
The primary challenge of defining a convolution operation over the hypergraph is how the embeddings of items are propagated. Referring to the spectral hypergraph convolution proposed in \cite{feng2019hypergraph}, we define our hypergraph convolution as:
\begin{equation}
\mathbf{x}_{i}^{(l+1)}=\sum_{j=1}^{N} \sum_{\epsilon=1}^{M} H_{i \epsilon} H_{j \epsilon} W_{\epsilon \epsilon} \mathbf{x}_{j}^{(l)}.
\end{equation}
Following the suggestions in \cite{wu2019simplifying}, we do not use nonlinear activation function and the convolution filter parameter matrix. For $W_{\epsilon\epsilon}$, we assign each hyperedge the same weight 1. The matrix form of Eq. (1) with row normalization is:
\begin{equation}
\mathbf{X}_{h}^{(l+1)}=\mathbf{D}^{-1} \mathbf{H} \mathbf{WB^{-1}} \mathbf{H}^{\mathrm{T}} \mathbf{X}_{h}^{(l)}.
\end{equation}
The hypergraph convolution can be viewed as a two-stage refinement performing `node-hyperedge-node' feature transformation upon hypergraph structure. The multiplication operation $\mathbf{H}^{\top}\mathbf{X}^{(l)}_{h}$ defines the information aggregation from nodes to hyperedges and then premultiplying $\mathbf{H}$ is viewed to aggregate information from hyperedges to nodes. \par
After passing $\mathbf{X}^{(0)}$ through $L$ hypergraph convolutional layer, we average the items embeddings obtained at each layer to get the final item embeddings $\mathbf{X}_{h}=\frac{1}{L+1}\sum_{l=0}^{L}\mathbf{X}^{(l)}_{h}$. Although this work mainly emphasizes the importance of the coherence of a session, the temporal information is also inevitable for better recommendation results. Position Embeddings is an effective technique which was introduced in Transformer \cite{vaswani2017attention} and has been applied in many situations for the memory of position information of items. In our method, we integrate the reversed position embeddings with the learned item representations by a learnable position matrix $\mathbf{P}_{r} = \left[\mathbf{p_{1}},\mathbf{p_{2}},\mathbf{p_{3}}, ...,\mathbf{p_{m}}\right]$, where $m$ is the length of the current session. The embedding of $t$-th item in session $s = [i_{s,1}, i_{s,2}, i_{s,3}, ..., i_{s,m}]$ is:
\begin{equation}
\mathbf{x}^{*}_{t}=\tanh \left(\mathbf{W}_{1}\left[\mathbf{x}_{t} \| \mathbf{p}_{m-i+1}\right]+\mathbf{b}\right),
\end{equation}
where $\mathbf{W}_{1}\in\mathbb{R}^{d \times 2d}$, and $b\in\mathbb{R}^{d}$ are learnable parameters.\par

Session embeddings can be represented by aggregating representation of items in that session. We follow the strategy used in SR-GNN \cite{wu2019session} to refine the embedding of session $s = [i_{s,1}, i_{s,2}, i_{s,3}, ..., i_{s,m}]$: 
\begin{equation}
\begin{aligned}
    \alpha_{t}=\mathbf{f}^{\top} \sigma\left(\mathbf{W}_{2} \mathbf{x}^{*}_{s}+\mathbf{W}_{3} \mathbf{x}^{*}_{t}+\mathbf{c}\right), \mathbf{\theta}_{h}=\sum_{t=1}^{m} \alpha_{t} \mathbf{x}_{t}
\end{aligned}
\end{equation}
where $\mathbf{x}^{*}_{s}$ is the embedding of session $s$ and here it is represented by averaging the embeddings of items it contains, which is $\mathbf{x}^{*}_{s} = \frac{1}{m}\sum_{t=1}^{m}\mathbf{x}_{m}$, and $\mathbf{x}^{*}_{t}$ is the embedding of the $t$-th item in session $s$. User's general interest embedding $\mathbf{\theta}_{h}$ across this session is represented by aggregating item embeddings through a soft-attention mechanism where items have different levels of priorities. $\mathbf{f}\in \mathbb{R}^{d}$, $\mathbf{W}_{2}\in\mathbb{R}^{d \times d}$ and $\mathbf{W}_{3}\in\mathbb{R}^{d \times d}$ are attention parameters used to learn the item weight $\alpha_{t}$.  Note that, following our motivation in Section I, we abandon the sequence modeling techniques like GRU units and self-attention used in other SBR models. The position embedding is the only temporal factor we use, and hence our model is very efficient and lightweight.
\subsubsection{Model Optimization and Recommendation Generation}
Given a session $s$, we compute scores $\hat{\mathbf{z}}$ for all the candidate items $i \in I$ by doing inner product between the item embedding $\mathbf{X}_{h}$ learned from hypergraph channel and $\mathbf{\theta}_{h}$:
\begin{equation}
\hat{\mathbf{z}}_{i}=\mathbf{\theta}_{h}^{T}\mathbf{x}_{i}.
\end{equation}
After that, a softmax function is applied to compute the probabilities of each item being the next one in the session:
\begin{equation}
\hat{\mathbf{y}}=\operatorname{softmax}(\hat{\mathbf{z}}).
\end{equation}

We formulate the learning objective as a cross entropy loss function, which has been extensively used in recommender systems and defined as:
\begin{equation}
\mathcal{L}_{r}=-\sum_{i=1}^{N} \mathbf{y}_{i} \log \left(\hat{\mathbf{y}}_{i}\right)+\left(1-\mathbf{y}_{i}\right) \log \left(1-\hat{\mathbf{y}}_{i}\right),
\end{equation}
where $\mathbf{y}$ is the one-hot encoding vector of the ground truth. For simplicity, we leave out the $L_{2}$ regularization terms.  By minimizing $\mathcal{L}_{r}$ with Adam, we can get high-quality session-based recommendations.

\subsection{Enhancing SBR with Self-Supervised Learning}
The hypergraph modeling empowers our model to achieve significant performance. However, we consider that the sparsity of session data might impede hypergraph modeling, which would result in a suboptimal recommendation performance. Inspired by the successful practices of self-supervised learning on simple graphs, we innovatively integrate self-supervised learning into the network to enhance hypergraph modeling. We first design another graph convolutional network based on the line graph of the session-induced hypergraph to generate self-supervision signals. Then by maximizing the mutual information between the session representations learned via the two channels through contrastive learning, the recommendation model can acquire more information and the recommendation performance can be improved. Since the two types of networks both are based on hypergraph, which can be seen as two channels for hypergraph modeling, we name our model as \textbf{DHCN} (Dual Channel Hypergraph Convolutional Networks).

\subsubsection{Line Graph Channel and Convolution}
The line graph channel encodes the line graph of the hypergraph. Fig. 1 shows how we transform the hypergraph into a line graph of it. The line graph can be seen as a simple graph which contains the cross-session information and depicts the connectivity of hyperedges.  As there are no item involved in the line graph channel, we first initialize the channel-specific session embeddings $\mathbf{\Theta}_{l}^{(0)}$ by looking up the items belonged to each session and then averaging the corresponding items embeddings in $\mathbf{X}^{(0)}$. An incidence matrix for $L(G)$ is defined as $\mathbf{A}\in\mathbb{R}^{M \times M}$ where $M$ is the number of nodes in the line graph and $A_{p, q}=W_{p, q}$ according to Definition 2. Let $\hat{\mathbf{A}}=\mathbf{A}+\mathbf{I}$ where $\mathbf{I}$ is an identity matrix. $\hat{\mathbf{D}}\in\mathbb{R}^{M \times M}$ is a diagonal degree matrix where $\hat{\mathbf{D}}_{p, p}=\sum_{q=1}^{m}\hat{\mathbf{A}}_{p, q}$. The line graph convolution is then defined as:
\begin{equation}
\mathbf{\Theta}^{(l+1)}_{l}=\mathbf{\hat{D}}^{-1} \mathbf{\hat{A}} \mathbf{\Theta}^{(l)}.
\end{equation}
 In each convolution, the sessions gather information from their neighbors. By doing so, the learned $\mathbf{\Theta}$ can capture the cross-session information. Likewise, we pass $\mathbf{\Theta}_{l}^{(0)}$ through $L$ graph convolutional layer, and then average the session embeddings obtained at each layer to get the final session embeddings $\mathbf{\Theta}_{l}=\frac{1}{L+1}\sum_{l=0}^{L}\mathbf{\Theta}^{(l)}_{l}$.

\subsubsection{Creating self-supervision signals.} So far, we learn two groups of channel-specific session embeddings via the two channels. Since each channel encodes a (hyper)graph that only depicts either of the item-level (intra-session) or the session-level (inter-session) structural information of the session-induced hypergraph, the two groups of embeddings know little about each other but can mutually complement. For each mini-batch including $n$ sessions in the training, there is a bijective mapping between the two groups of session embeddings. Straightforwardly, the two groups can be the ground-truth of each other for self-supervised learning, and this one-to-one mapping is seen as the label augmentation. If two session embeddings both denote the same session in two views, we label this pair as the ground-truth, otherwise we label it as the negative.\\
\subsubsection{Contrastive learning.} Following \cite{velickovic2019deep,bachman2019learning}, we regard the two channels in DHCN as two views characterizing different aspects of sessions. We then contrast the two groups of session embeddings learned via the two views. We use a noise-contrastive type objective with a standard binary cross-entropy (BCE) loss between the samples from the ground-truth (positive) and the corrupted samples (negative) as our learning objective and defined it as:
\begin{equation}
\mathcal{L}_{s}=-\log\sigma(f_{\mathrm{D}}(\theta^{h}_{i}, \theta^{l}_{i}))-\log\sigma(1- f_{\mathrm{D}}(\tilde\theta^{h}_{i}, \theta^{l}_{i})),
\end{equation}
where $\tilde\theta^{h}_{i}$ (or $\tilde\theta^{l}_{i}$) is the negative sample obtained by corrupting $\Theta_{h}$ ($\Theta_{l}$) with row-wise and column-wise shuffling, and $f_{\mathrm{D}}(\cdot): \mathbb{R}^{d} \times \mathbb{R}^{d} \longmapsto \mathbb{R}$ is the discriminator function that takes two vectors as the input and then scores the agreement between them. We simply implement the discriminator as the dot product between two vectors. This learning objective is explained as maximizing the mutual information between the session embeddings learned in different views \cite{velickovic2019deep}. By doing so, they can acquire information from each other to improve their own performance in item/session feature extraction through the convolution operations. Particularly, those sessions that only include a few items can leverage the cross-session information to refine their embeddings. 
\par
Finally, we unify the recommendation task and this self-supervised task into a \textit{primary\&auxiliary} learning framework, where the former is the primary task and the latter is the auxiliary task. Formally, the joint learning objective is defined as:
\begin{equation}
\mathcal{L}=\mathcal{L}_{r}+\beta\mathcal{L}_{s},
\end{equation}
where $\beta$ controls the magnitude of the self-supervised task.

\section{Experiments}

\subsection{Experimental Settings}
\subsubsection{Datasets.}
We evaluate our model on two real-world benchmark datasets: \textit{Tmall}\footnote{https://tianchi.aliyun.com/dataset/dataDetail?dataId=42}, \textit{Nowplaying}\footnote{http://dbis-nowplaying.uibk.ac.at/\#nowplaying} and \textit{Diginetica}\footnote{http://cikm2016.cs.iupui.edu/cikm-cup/}. Tmall dataset
comes from IJCAI-15 competition, which contains anonymized
user’s shopping logs on Tmall online shopping platform. Nowplaying dataset describes the music listening behavior of users.
For both datasets, we follow~\cite{wu2019session,li2017neural} to remove all sessions containing only one item and also remove items appearing less than five times. To evaluate our model, we split both datasets into training/test sets,  following the settings in ~\cite{wu2019session,li2017neural,wang2020global}. 
Then, we augment and label the dataset by using a sequence splitting method, which generates multiple labeled sequences with the corresponding labels $([i_{s,1}], i_{s,2}), ([i_{s,1}, i_{s,2}], i_{s,3}), ...,
([i_{s,1}, i_{s,2}, ..., i_{s,m-1}], i_{s,m})$ for every session $s = [i_{s,1}, i_{s,2}, i_{s,3}, ..., i_{s,m}]$. Note that the label of each sequence is the last click item in it. 
The statistics of the datasets are presented in Table 1.
\begin{table}[h]
	\renewcommand\arraystretch{1.0}
	\label{Table:1}
	\begin{center}
		\begin{tabular}{ccccc}
			\hline
			Dataset & Tmall & Nowplaying & Diginetica \\ \hline
			training sessions & 351,268 & 825,304 & 719,470\\
			test sessions & 25,898 & 89,824 & 60,858 \\
			\# of items & 40,728 & 60,417 & 43,097 \\
			average lengths & 6.69 & 7.42 & 5.12 \\
			\hline
		\end{tabular}
	\end{center}
	\caption{Dataset Statistics}
\end{table}


\begin{table*}[tp]
		\small
	\label{Table:2}
	\renewcommand\arraystretch{1.1}
	\begin{center}
		{
		\begin{threeparttable}{
			\begin{tabular}{*{13}{c}}
				\toprule
				\multirow{2}{*}{Method} &
				\multicolumn{4}{c}{Tmall} & \multicolumn{4}{c}{Nowplaying} & \multicolumn{4}{c}{Diginetica} \cr
				\cmidrule(lr){2-5}\cmidrule(lr){6-9}\cmidrule(lr){10-13} & P@10 & M@10 & P@20 & M@20 & P@10 & M@10 & P@20 & M@20 & P@10 & M@10 & P@20 & M@20  \\ \hline

				Item-KNN & 6.65 & 3.11 & 9.15 & 3.31 & 10.96 & 4.55 & 15.94 & 4.91 & 25.07 & 10.77 & 35.75 & 11.57 \\
				
				FPMC &13.10 & 7.12 & 16.06 & 7.32 & 5.28 & 2.68 & 7.36 &2.82  & 15.43 & 6.20  & 26.53 & 6.95 \\
				
				GRU4REC  & 9.47 & 5.78 & 10.93 & 5.89 & 6.74 & 4.40 & 7.92 & 4.48 & 17.93 & 7.33  & 29.45 & 8.33 \\
				
				NARM   & 19.17  & 10.42 & 23.30 & 10.70 & 13.6  & 6.62 & 18.59 & 6.93 & 35.44 & 15.13 & 49.70 & 16.17 \\
				
				STAMP  & 22.63 & 13.12  &26.47 &13.36 & 13.22 & 6.57 & 17.66 & 6.88 & 33.98 & 14.26 & 45.64 & 14.32 \\
				
				SR-GNN & 23.41  &13.45  & 27.57 & 13.72 & 14.17 & 7.15 & 18.87 & 7.47 & 36.86 & 15.52 & 50.73 & 17.59 \\
				
				FGNN   & 20.67 & 10.07 & 25.24 & 10.39 & 13.89 & 6.8 & 18.78 & 7.15 & 37.72 & 15.95 &  50.58& 16.84 \\
				\hline 
				DHCN & \textbf{25.14}\tnote{*} & \textbf{13.91}\tnote{*} & \textbf{30.43}\tnote{*} & \textbf{14.26}\tnote{*} & \textbf{17.22}& \textbf{7.78}&\textbf{23.03} & \textbf{8.18}& \textbf{39.87} & \textbf{17.53}& \textbf{53.18} & \textbf{18.44} \\
				$S^{2}$-DHCN & \textbf{26.22} & \textbf{14.60} & \textbf{31.42} & \textbf{15.05} &\textbf{17.35} &\textbf{7.87}&\textbf{23.50}&\textbf{8.18}&\textbf{40.21} & \textbf{17.59}& \textbf{53.66}& \textbf{18.51}\\ \hline
				Improv. (\%) & 10.71 & 7.87 & 12.25 & 8.84 & 18.32 &9.15 & 19.70&8.68 & 6.19& 9.32& 5.46& 4.97 \\
				\bottomrule
		\end{tabular}}
		\begin{tablenotes}
               {\footnotesize
               \item[*] The reported results on Tmall are not the best here. Refer to the ablation study in Section 4 for the best results.}
               \end{tablenotes}
        
		\end{threeparttable}}
	\end{center}
	\caption{Performances of all comparison methods on three datasets.}
	
	\vspace{-10pt}

\end{table*}

\subsubsection{Baseline Methods.}
We compare DHCN with the following representative methods:
\begin{itemize}
	\item \textbf{Item-KNN}\cite{sarwar2001item} recommends items similar to the previously clicked item in the session, where the cosine similarity between the
	vector of sessions is used.
	\item \textbf{FPMC} \cite{rendle2010factorizing} is a sequential method based on Markov Chain.
	\item \textbf{GRU4REC} \cite{hidasi2015session} utilizes a session-parallel mini-batch training process and adopts ranking-based loss functions to model user sequences.
	\item \textbf{NARM} \cite{li2017neural}: is a RNN-based model that models the sequential behavior to generate the recommendations.
	\item \textbf{STAMP} \cite{liu2018stamp}: employs the self-attention mechanism to enhance session-based recommendation.
	\item \textbf{SR-GNN} \cite{wu2019session}: applies a gated graph convolutional layer  to learn item transitions.
	\item \textbf{FGNN} \cite{qiu2019rethinking}: formulates the next item recommendation within the session as a graph classification problem.
\end{itemize}

\subsubsection{Evaluation Metrics.}
Following \cite{wu2019session,liu2018stamp}, we use P@K (Precision) and MRR@K (Mean Reciprocal Rank) to evaluate the recommendation results.

\subsubsection{Hyper-parameters Settings.}
For the general setting, the embedding size is 100, the batch size for mini-batch is 100, and the $L_2$ regularization is  $10^{-5}$. For DHCN, an initial learning rate 0.001 is used. The number of layers is different in different datasets. For \textit{Nowplaying} and \textit{Diginetica}, a three-layer setting is the best, while for \textit{Tmall}, one-layer setting achieves the best performance. For the baseline models, we refer to their best parameter setups reported in the original papers and directly report their results if available, since we use the same datasets and evaluation settings.

\subsection{Experimental Results}

\subsubsection{Overall Performance.}
The experimental results of overall performance are reported in Table 2, and we highlight the best results of each column in boldface. Two variants of DHCN are evaluated, and \textbf{$S^{2}$-DHCN} denotes the self-supervised version. The improvements are calculated by using the difference between the performance of $S^{2}$-DHCN and the best baseline to divide the performance of the latter. Analyzing the results in Table 2, we can draw the following conclusions. 
\begin{itemize}
\item The recently proposed models that consider the sequential dependency in the sessions (i.e., GRU4REC, NARM, STAMP, SR-GNN and DHCN) significantly outperform the traditional  models that do not (i.e., FPMC).
This demonstrates the importance of sequential effects for session-based recommendation. Furthermore, the fact that GRU4REC, NARM, STAMP, SR-GNN and DHCN all employ the deep learning technique confirms its key role in session-based recommendation models.
\item For the baseline models based on deep recurrent neural structure (e.g., RNN, LSTM and GRU), NARM obtains higher accuracy in all settings. This is because that GRU4REC only takes the sequential behavior into account and may have difficulty in dealing with the shift of user preference. By contrast, NARM and STAMP uses recurrent units to encode user behaviors and exerts an attention mechanism over the items in a session, improving the recommendation results by a large margin. The superior performance of NARM and STAMP proves that assigning various importance value on different items within the session help formulate user intent more accurately. Besides, STAMP outperforms NARM by incorporating the short-term priority over the last item in a session, further demonstrating that directly using RNN to learn user representations may lead to recommendation bias but this can be avoided by replacing it with the attention mechanism. 
\item The GNNs-based models: SR-GNN and FGNN outperform RNNs-based models. The improvements can be owed to the great capacity of graph neural networks. However, the improvements are also trivial compared with the improvements brought by DHCN. 
 
\item Our proposed DHCN shows overwhelming superiority over all the baselines on all datasets. 
Compared with SR-GNN and FGNN, our model has two advantages: (1) It uses hypergraph to capture the beyond pairwise relations. By modeling each hyperedge as a clique whose items are fully connected, the connections between distant items can be exploited. (2) Also, our DHCN is lightweight than the SR-GNN and FGNN because we use very limited parameters in hypergraph convolution of the two channels, showing the efficicency of DHCN. 
\item Although not as considerable as those brought by hypergraph modeling, the improvements brought by self-supervised learning are still decent. In particular, on the two datasets which have shorter average length of sessions, self-supervised learning plays a more important role, which is line with our assumption that the sparsity of session data might hinder the benefits of hypergraph modeling, and maximizing mutual information between the two views in DHCN could address it. 
\end{itemize}
 \begin{figure}[t]
	\centering
	\includegraphics[width=0.49\textwidth]{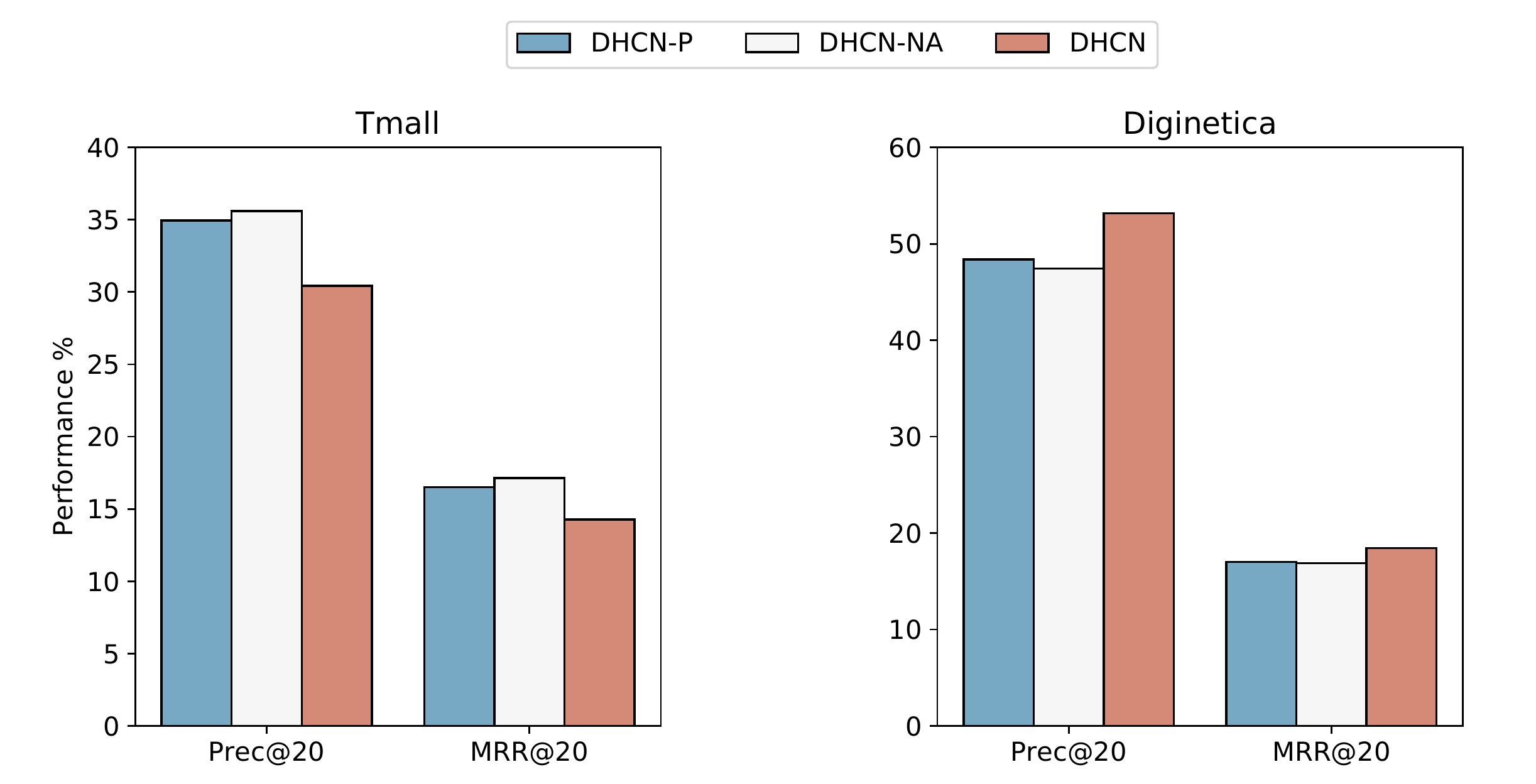}
	\caption{Contribution of each component.}
	\label{figure.3}
\end{figure}

\subsubsection{Ablation Study.}
The overwhelming superiority of DHCN shown in the last section can be seen as the result of the joint effect of hypergraph modeling, and temporal factor exploitation. To investigate the contributions of each module in DHCN, we develop two variants of DHCN: \textbf{DHCN-P} and \textbf{DHCN-NA}.  DHCN-P represents the version without the reversed position embeddings, and DHCN-NA represents the version without the soft attention mechanism. We compare them with the full DHCN on \textit{Tmall} and \textit{Diginetica}.
\par
As can be observed in Figure 2, the contributions of each component are different on the two datasets. For Tmall, to our surprise, when removing the reversed position embeddings or soft attention, the simplified version achieves a performance increase on both metrics and the performance is even better than that of the the full version. Considering that the Tmall dataset is collected in a real e-commerce situation, this finding, to some degree, validates our assumption that coherence may be more important than strict order modeling. By contrast, in Diginetica, the reversed position embeddings and soft attention are beneficial. When removing reversed position embedding or soft attention, there is a performance drop on both metrics. Soft attention contributes more on Diginetica, demonstraing the importance of different priorities of items when generating recommendation. 

\subsubsection{Impact of Model Depth.}
To study the impacts of hypergraph convolutional network's depth in session-based recommendation, we range the numbers of layers of the network within \{1, 2, 3, 4, 5\}. According to the results presented in Figure 3, DHCN is not very sensitive to the number of layers on Diginetica and a three-layer setting is the best. However, on Tmall, a one-layer network achieves the best performance. Besides, with the number of layer increases, the performance on MRR@20 drops. The possible cause could be the increasingly over-smoothed representations of items.  

\begin{figure}[t]
	\centering
	\includegraphics[width=0.49\textwidth]{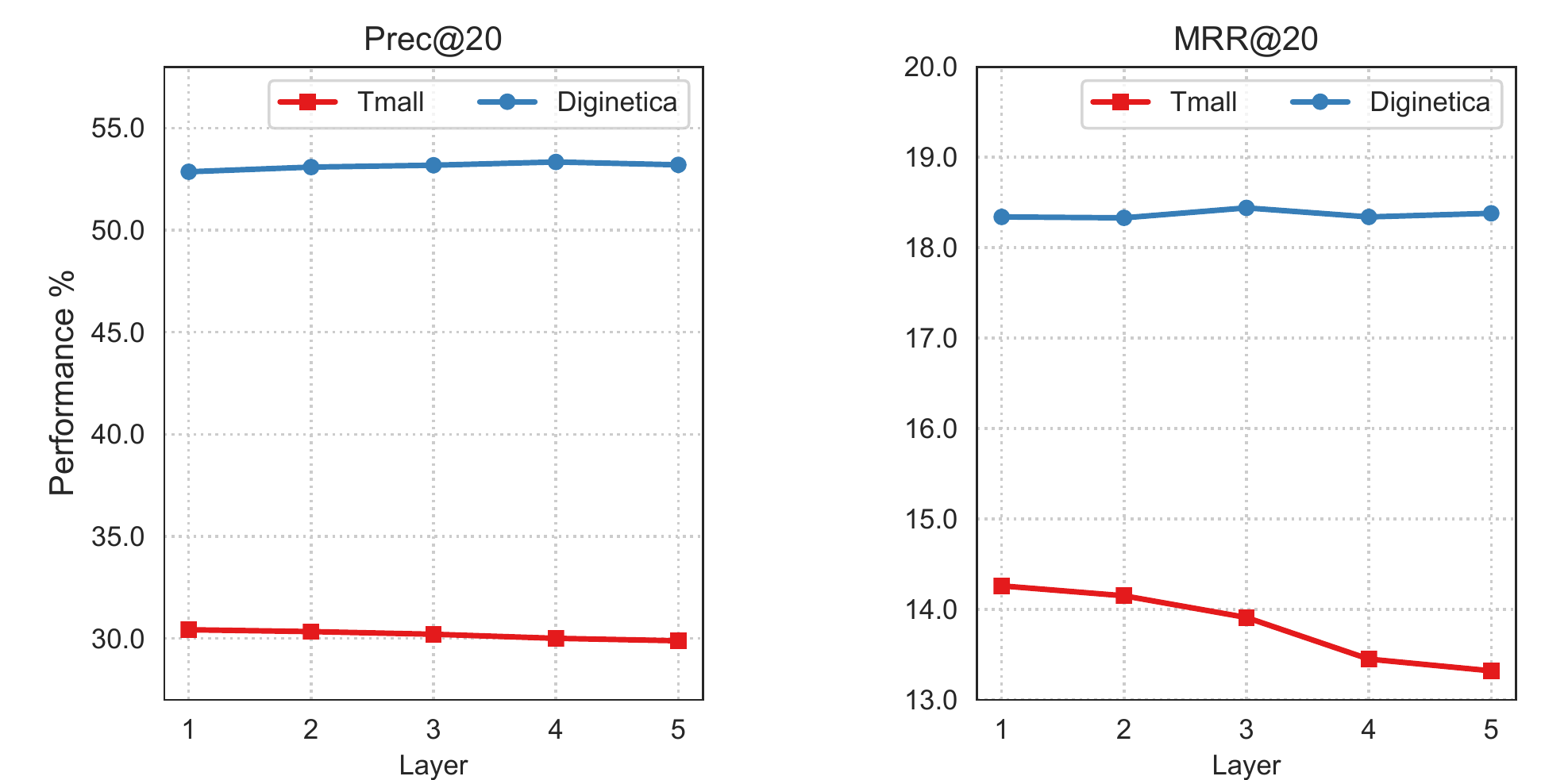}
	\caption{The impacts of the number of layer.}
	\label{figure.3}
\end{figure}

\begin{figure}[t]
	\centering
	\includegraphics[width=0.49\textwidth]{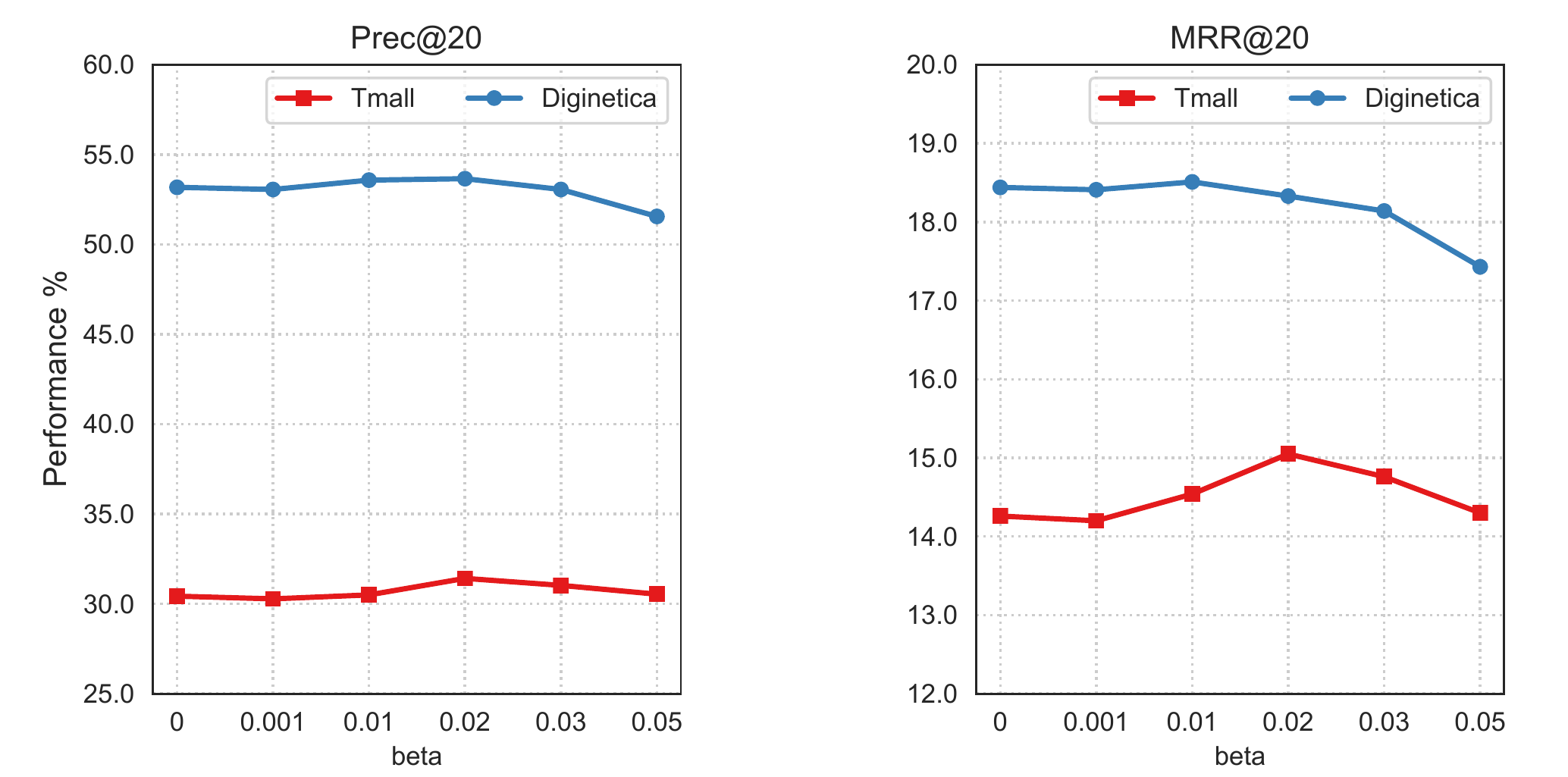}
	\caption{The impact of the magnitude of self-supervised learning.}
	\label{figure.4}
\end{figure}

\subsubsection{Impact of Self-Supervised Learning.}
We introduce a hyper-parameter $\beta$ to $S^{2}$-DHCN to control the magnitude of self-supervised learning. To investigate the influence of the self-supervised task based on two-view contrastive learning, we report the performance of $S^{2}$-DHCN with a set of representative $\beta$ values \{0.001, 0.01, 0.02, 0.03, 0.05\}. According to the results presented in Figure 4, recommendation task achieves decent gains when jointly optimized with the self-supervised task. 
For both datasets, learning with smaller $\beta$ values can boost both Prec@20 and MRR@20, and with the increase of $\beta$, the performance declines. We think it is led due to the gradient conflicts between the two tasks. Besides, with lager $beta$, performance declines obviously on MRR@20, which means that in some cases, it is important to make a trade-off between the hit ratio and item ranks when choosing the value of $\beta$.

\section{Conclusion}
\label{st:conc}
Existing GNNs-based SBR models regard the item transitions as pairwise relations, which cannot capture the ubiquitous high-order correlations among items. In this paper, we propose a dual channel hypergraph convolutional network for SBR to address this problem, Moreover, to further enhance the network, we innovatively integrate self-supervised into the training of the network. Extensive empirical studies demonstrate the overwhelming superiority of our model, and the ablation study validates the effectiveness and rationale of the hypergraph convolution and self-supervised learning.
\section*{Acknowledgments}
This work was supported by ARC Discovery Project (GrantNo.DP190101985, DP170103954).

\bibliography{ref}
\end{document}